\documentclass[12pt,tightenlines,floats,showpacs,nofootinbib,amsmath,amssymb,aps,prd]{revtex4-2}

\usepackage{tikz}
\usepackage{graphicx,verbatim}
\usepackage{amsmath}
\usepackage{amsfonts}
\usepackage{amssymb}
\usepackage{psfrag}
\usepackage{accents}
\usepackage{mathrsfs}
\usepackage{enumitem}
\usepackage{hyperref}

\newcommand*{\scrip}{\ensuremath{\mathscr{I}^{+}}}

\def\Deltap{\Delta^{\!{}^+}}
\usepackage[colorinlistoftodos]{todonotes}
\usepackage{epstopdf}

\def\be{\begin{equation}}
\def\ee{\end{equation}}
\def\ba{\begin{eqnarray}}
\def\ea{\end{eqnarray}}

\newcommand{\pb}[1]{\hbox{\lower0.5ex\hbox{${}_{\leftarrow}$}}\kern-1.9ex{#1}}

\def\h{\hat}
\def\b{\bar}

\def\={\,\hat{=}\,}
\def\f{\frac}

\def\Lie{\mathcal{L}}
\def\ps{\boldsymbol{\Gamma}_{\rm cov}}
\def\psR{\boldsymbol{\Gamma}_{\!\scriptscriptstyle{\cal R}}}
\def\psRh{\boldsymbol{\Gamma}_{\!\scriptscriptstyle{\cal {\hat R}}}}

\def\bfomega{\boldsymbol{\omega}}

\def\bfomegaR{\boldsymbol{\omega}_{\scriptscriptstyle{\cal R}}}
\def\bfomegaRh{\boldsymbol{\omega}_{\scriptscriptstyle{\cal {\hat R}}}}

\def\bM{\bar{M}}
\def\bg{\bar{g}}
\def\bR{\bar{R}}
\def\bC{\bar{C}}

\def\bD{\bar{D}}
\def\bq{\bar{q}}

\def\bS{\bar{S}}
\def\bomega{\bar\omega}
\def\hM{\hat{M}}
\def\hg{\hat{g}}

\def\hD{\hat{D}}
\def\hn{\hat{n}}
\def\hq{\hat{q}}
\def\hS{\hat{S}}

\def\hN{\hat{N}}
\def\bk{\bar{k}}
\def\ko{\mathring{k}}
\def\bj{\bar{\jmath}}

\def\G{\mathfrak{G}}

\def\B{\mathfrak{B}} 

\def\J{\mathfrak{J}}

\def\qo{\mathring{q}}

\def\vo{\mathring{v}}

\def\betao{\mathring{\beta}}
\def\chio{\mathring{\chi}}
\def\psio{\mathring{\psi}}
\def\alphao{\mathring\alpha}

\def\ello{\mathring{\ell}}
\def\no{\mathring{n}}

\def\uo{\mathring{u}}

\def\Do{\mathring{D}}

\def\epsilono{\mathring{\epsilon}}

\def\WIH{\mathfrak{h}}

\def\psio{\mathring{\psi}}

\def\R{\mathcal{R}}

\def\Rh{\hat{\mathcal{R}}}

\begin{document}

\title[]{Horizons and Null Infinity: A Fugue in 4 voices}
 
\author{Abhay Ashtekar}
\email{ashtekar.gravity@gmail.com}
\affiliation{Institute for Gravitation and the Cosmos, Pennsylvania State 
University, University Park, PA 16802, USA}
\affiliation{Perimeter Institute for Theoretical Physics, 31 Caroline St N, Waterloo, ON N2L 2Y5, Canada}

\author{Simone Speziale}
\email{simone.speziale@cpt.univ-mrs.fr}
\affiliation{Aix Marseille Univ., Univ. de Toulon, CNRS, CPT, UMR 7332, 13288 Marseille, France}

\begin{abstract}

Black hole horizons in equilibrium and null infinity of asymptotically flat space-times are null 3-manifolds but have very different physical connotations. We first show that they share a large number of geometric properties, making them both weakly isolated horizons.  We then use this new unified perspective to unravel the origin of the drastic differences in the physics they contain. Interestingly, the themes are woven together in a manner reminiscent of voices in a fugue.

\end{abstract}

\maketitle

\textbf{Introduction:}\, Two types of boundaries commonly arise in general relativity: future horizons $\Deltap$ of black holes in equilibrium and Penrose's future null infinity $\scrip$. While it is obvious that both these boundaries are null 3-surfaces, it is not well appreciated that they share other, more subtle geometric structures as well. In particular each of them inherits an intrinsically defined derivative operator with certain properties that ensure that they are both \emph{Weakly Isolated Horizons} (WIHs) \cite{aass1,afk,abl1}. At first this appears surprising --even impossible-- because while WIHs are generally thought of as representing black hole boundaries in strong field regions, $\scrip$ lies in  asymptotic, near-flat regions. More importantly, WIHs are often associated with \emph{`stationary'} black holes since there is no flux of gravitational waves (or matter fields) across them, while $\scrip$ is the quintessential arena to calculate the fluxes of energy and momenta carried by gravitational waves (as well as other massless fields). The purpose of this Letter is to summarize recent work \cite{aass1} that compares and contrasts their geometry and physics, and resolves this apparent tension.

As we recall below, a  WIH horizon $\WIH$ can be defined purely in geometrical terms, \emph{without reference} to any field equations. Nonetheless, WIHs carry a universal structure which is very similar to that of $\scrip$. As a result, the symmetry group $\G$ of WIHs $\WIH$ is closely related to the BMS group \cite{akkl1}. Second, the intrinsic (degenerate) metric $\bq_{ab}$ is non-dynamical on $\WIH$ just as it at $\scrip$. The third common element seems counter intuitive at first: \emph{the intrinsic connection $\bD$ on a general $\WIH$ is dynamical} just as it is at $\scrip$. This important point is often overlooked in the literature since WIHs have been often regarded as `stationary boundaries'. 

Both black hole horizons $\Deltap$ and null infinity $\scrip$ are examples of geometric WIHs $\WIH$. However, while  $\Deltap$ is a WIH with respect to the \emph{physical} metric $g_{ab}$ that satisfies Einstein's equations,  $\scrip$ is a WIH with respect to the rescaled  metric $\hg_{ab}$ used in the Penrose completion that satisfies \emph{conformal} Einstein's equations. Einstein's equations imply that the time dependence of the connection $D$ on $\Deltap$ is extremely constrained: $D$ is completely determined on the entire 3-manifold $\Deltap$ by `data' that can be specified on a 2-dimensional cross-section \cite{abl1}. Therefore, in spite of its time dependence, $D$ does not carry \emph{any} 3-d degrees of freedom. Physically this means that `radiative modes' are absent on $\Deltap$; the connection $D$ on $\Deltap$ encodes only `Coulombic information'.  The situation is just the opposite at $\scrip$: conformal Einstein's equations imply that the intrinsic connection $\hD$ captures precisely the two radiative modes of the gravitational field \cite{aa-radiativemodes}; it carries no `Coulombic information'! We will see  that this striking difference emerges from the \emph{same} evolution equation, Eq~(\ref{Ddot1}), satisfied by the connection on any WIH $\WIH$. It is remarkable that both the close geometrical similarities and the diametrically opposite physics emerges from the common WIH framework. 

This structure is also subtle. For example, one may envisage using isolated horizons (IHs) \cite{Ashtekar_2000} in place of WIHs. This strengthening of conditions may seem innocuous since the notion of IHs is also considerably weaker than that of Killing horizons. (For example, the Robinson-Trautman and Kastor-Traschen solutions admit IHs but no Killing horizons \cite{ptc,Kastor_1993}.) But passage from WIHs to IHs is already unacceptably strong for our purposes because while $\scrip$ of any asymptotically flat space-time is a WIH, it would be an IH only if the Bondi news vanishes, i.e., there is no radiation across $\scrip$!

As discussed in \cite{aass1} all our considerations apply also to past horizons and past null infinity and, in presence of a positive $\Lambda$, also to cosmological horizons $\Delta$ which may not be associated with black holes. In this brief report we focus on $\Delta^+$ and $\scrip$ because physically they are the most interesting cases. Also, for simplicity we assume that the physical metric satisfies Einstein's equations in a neighborhood of $\Deltap$ and $\scrip$; one can allow appropriate sources.
\\

\begin{table}
\begin{center}
\footnotesize
\begin{tabular}{|l|c|c|c|}
\hline
\quad Notions\quad   &\,\,\, General spacetimes\,\,\, & \,\,\,Physical spacetimes\,\,\, &\,\,\, Conformal Completions\,\,\,\\ \hline\hline
Field Eqs &  None & Einstein's Eqs & Conformal Einstein's Eqs \\ \hline
4-Manifolds \&  metrics thereon  & $\b{M}, \b{g}_{ab}$ & ${M}, {g}_{ab}$ & $\h{M}, \h{g}_{ab}$ \\ \hline
Horizons & $\WIH$ & $ \Delta^+$ & $\scrip$ \\ \hline
Induced metrics & $\b{q}_{ab}$&$q_{ab}$&$\h{q}_{ab}$ \\ \hline
Null normals &$\b{k}^a$&$\ell^a$&$\hn^a$ \\ \hline
Dual 1-forms & $\bar\jmath_a$ &$n_a$ &$\hat\ell_a$ \\\hline
Intrinsic derivative operators & $ \b{D}$&$D$&$ \hD$\\\hline
Symmetry vector fields &$\xi^a$&$\xi^a$&$\xi^a$ \\\hline
Symmetry groups & $\G=\B \ltimes \mathfrak{D}$ &$\G$ & $\B$ \\\hline
Local regions & &$\R$ & $\h\R$ \\\hline
Covariant and local phase spaces & & $\ps,\ \psR$& $ \ps,\ \psRh$ \\\hline
\end{tabular}
\caption{\emph{Symbols associated with WIHs and their meaning}}
\label{tab1}
\end{center}
\end{table}

\textbf{Geometric WIHs} $\WIH$:\, Fix a 4-manifold $\bM$ with a -,+,+,+ metric $\bg_{ab}$. Consider a null, 3-dimensional sub-manifold $\WIH$ of $\bM$ with topology $\mathbb{S}^2\times \mathbb{R}$. Recall that it is said to be a Non-Expanding Horizon (NEH) if \cite{akkl1}:\\
\indent (i) Every null normal ${\bar{k}}^a$ to $\WIH$ is expansion free:\,\, $\theta_{({\bar{k}})} \= 0$ where $\=$  \emph{stands for equality at points of\, $\WIH$}. For definiteness, we will assume that the null normals ${\bar{k}}^a$ are all future pointing;\, and,\\
\indent (ii) The Ricci tensor $\bR_a{}^b$ of $\bg_{ab}$ satisfies $\bR_a{}^b {\bar{k}}^a\, \= \,\alpha {\bar{k}}^b$ for some function $\alpha$.\vskip0.1cm 
There is extensive literature on NEHs (for reviews, see e.g., \cite{akrev,boothrev,gjrev,jaramillorev}) and their key  properties are summarized in \cite{aass1,akkl1}. In particular, the 4-metric $\bg_{ab}$ induces on $\WIH$, an intrinsic (degenerate) metric $\bq_{ab}$ and provides a family of null normals $\bk^a$ that are tangent to \emph{affinely parametrized} null geodesics. Any two of these normals are related by $\bk^{\prime\,a} \= f\, \bk^a$ where the positive function $f$ satisfies $\Lie_{\bk} f \=0$. The derivative operator $\b\nabla$ defined by $\bg_{ab}$ also induces a unique derivative operator $\bD$ on $\WIH$ with the following properties:
\be \label{bD} \bD_a \bq_{bc} \=0, \quad {\rm and}\quad \bD_a \bk^b \= \bomega_{a} \bk^b, \qquad {\hbox{\rm where $\bomega_{a}$ satisfies}}\quad
\bomega_{a} \bk^a \=0 \quad {\rm and}\quad \Lie_{\bk} \bomega_{a} \=0\, . \ee
($\b\omega_a$ depends on the choice of the null normal $\bk^a$ but we will suppress this dependence for notational simplicity.) One can eliminate the functional rescaling freedom in $\bk^a$ by requiring that the rotation 1-form $\bomega_a$ be divergence free: One is then left with a small class $[\b{k}]$ in which any two null normals are related by a $\bk^{\prime\,a} = c \bk^a$, for some positive \emph{constant} $c$ \cite{abl1}.

An NEH $\WIH$ equipped with such an equivalence class $[\bk^a]$ of null normals is said to be a \emph{Weakly Isolated Horizon}. (Our  WIHs are called \emph{extremal} because the acceleration of preferred null normals $[\bk^a]$ vanishes.) 
Eq.~(\ref{bD}) implies that the intrinsic metric $\bq_{ab}$ is time independent; $\Lie_{\bk}\, \bq_{ab}\=0$. However, on generic WIHs, the connection $\bD$ is time \emph{dependent}.\goodbreak \noindent It is easy to verify that if a 1-form $\b{f}_a$ is `horizontal', i.e., satisfies ${\bk^a} \b{f}_a =0$, then $\dot{\bD}_a \b{f}_b := (\Lie_{\bk} \bD_a - \bD_a \Lie_{\bk})\,\b{f}_b =0$. Therefore the `time derivative'  $\dot{\bD}$ is completely determined by  $\dot{\bD}_a \bj_b$ for any 1-form $\bj_a$ satisfying $\bj_a \bk^a =-1$. Since $\bR_{a}{}^{b}{\bar{k}}^a \= \alpha {\bar{k}}^b$ on any WIH $\WIH$, it follows that this action is given by \cite{aass1,abl1}
\be \label{Ddot1}\dot{\bD}_a \bj_b\, \= \bD_a\bomega_b + \bomega_a \bomega_b + {\bar{k}}^c\, \bC_{c\,{\pb{ab}}}{}^d\, \bj_d + \textstyle{\f{1}{2}}\,\big(\bS_{\pb{ab}} + (\alpha\,- \textstyle{\f{1}{6}}\b{R})\, \b{q}_{ab}\big)\,  \ee
where ${\bC}_{abc}{}^d$ and ${\bS}_{ab} = \b{R}_{ab} - \f{1}{6} \b{R}\, \b{g}_{ab}$ are the Weyl and Schouten tensors of $\bg_{ab}$, and the under arrow indices that indices $a,b$ are pulled back to $\WIH$. We will see that the drastic difference in the physics of black hole horizons $\Deltap$ and null infinity $\scrip$ can be directly traced back to Eq. (\ref{Ddot1}). On a generic WIH, none of the terms on the right side are zero, whence \emph{$\bD$ has genuine time dependence}. By contrast, $\dot{\bD} \=0$ on \emph{isolated} horizons (IHs)  by their very definition \cite{Ashtekar_2000}; this is why IHs are much more restrictive. In particular, while one can always choose a null normal on an NEH to endow it with the structure of a WIH, generically there is no null normal that can endow it with an IH structure.
  
Neither the metric $\bq_{ab}$ nor the connection $\bD$ on $\WIH$ are universal -- they change, for example, as one passes from the Schwarzschild WIH to the Kerr WIH. However, because of the $\mathbb{S}^2\times\mathbb{R}$ topology, each WIH admits a 3-parameter family of \emph{unit, round} 2-sphere metrics $\qo_{ab}$ that are conformally related to its $\bq_{ab}$: $\qo_{ab} \= \psio^2 \bq_{ab}$. While the conformal factors $\psio$ relating the metrics $\qo_{ab}$ and $\bq_{ab}$ vary from one WIH to another, the \emph{relative} conformal factors $\alphao$ between any two round metrics \emph{is universal}:
\be \label{trans} \qo^\prime_{ab} \= \alphao^2 \qo_{ab}\quad {\hbox{\rm where $\alphao$ satisfies}} \quad
\Do^2 \ln \alphao + 1\, \=\, \alpha_o^{-2}.\ee
The general solution to (\ref{trans}) is a linear combination of the first 4 spherical harmonics of $\qo_{ab}$: $\alpha_o^{-1} = \alpha_o + {\sum_m} \alpha_m Y_{1,m}$ where the coefficients are constrained to satisfy $- \alpha_o^2 + \sum_m |\alpha_m|^2 = -1$. Motivated by geometric considerations related to multipoles, one introduces a parallel rescaling of the vector fields $[{\bar{k}}^a]$ and sets $[\ko^a] \= [\psio^{-1}{\bar{k}}^a]$\,\,\cite{akkl1}. Then \emph{every} WIH is naturally equipped with a 3-parameter family of pairs $(\qo_{ab},\, [\ko^a])$ related by:\, $(\qo^\prime_{ab},\, [\ko^{\prime\,a}])\, \=\, (\alphao^2\, \qo_{ab},\, \alphao^{-1}\, [\ko^a])$. The relation is the same for all WIHs. Thus, the pairs capture the `kinematical' structure \emph{that is universal}, leaving out fields such as $\bq_{ab}, [{\bar{k}}^a], \bomega_a$ that can carry physical information that varies from one WIH to another. 
	
The WIH symmetry group $\G$ is the subgroup of the diffeomorphism group of $\WIH$ that preserves this universal structure. It is generated by vector fields $\xi^a$ satisfying 
\be \label{symmetry} \Lie_\xi\, \qo_{ab} \= 2\betao\, \qo_{ab} \qquad {\rm and} \qquad \Lie_\xi {\ko}^a \= - (\betao +\mathring{\varpi} ){\ko}^a\ee
where $\mathring{\varpi}$ is a constant and $\betao$ satisfies $\big(\Do^2 +2\big)\, \betao =0$;\, $\betao$  is a linear combination of the first three spherical harmonics of $\qo_{ab}$. The BMS vector fields $\xi^a$ at $\scrip$ satisfy the same equation (\ref{symmetry}) but with $\mathring{\varpi} =0$. This difference can be directly traced-back to the fact that while $\WIH$  is equipped only with an equivalence class $[\bk^a]$ of null normals (where any two are related by a constant rescaling), in any one conformal completion $\scrip$ is equipped with a canonical null normal $\hn^a$. Eq. (\ref{symmetry}) implies that $\G$ is a semi-direct product, $\G = \B \ltimes \mathfrak{D}$, of the BMS group $\B$ with the 1-dimensional group $\mathfrak{D}$ of dilations, with $\B$ serving as the normal subgroup and $\mathfrak{D}$, the quotient of $\G$ by $\B$.  An example of a dilation vector field is\, $d^a = \mathring{\varpi}\, \vo\, \ko^a = \mathring{\varpi}\, \b{v}\, \bk^a$ (where $\vo$ and $\b{v}$ are affine parameters of $\ko^a$ and $\bk^a$ respectively). Its action rescales $\ko^a$ (and $\bk^a$) by a constant, leaving $\qo_{ab}$ (and $\bq_{ab}$) untouched.  Could we perhaps have strengthened the boundary conditions defining WIHs $\WIH$ to eliminate this dilation and reduce $\G$ to $\B$? The answer is in the negative: On the Schwarzschild horizon, for example, the time translation is realized as the dilation in the Lie algebra of $\G$ (while it is realized as a supertranslation at $\scrip$). $\G$ has a rich structure that has been spelled out in \cite{akkl1}, and is a subgroup of the symmetry group of a general null hypersurface studied in \cite{cfp} (see also \cite{Freidel:2021yqe,Adami:2021nnf,Odak:2023pga,Chandrasekaran:2023vzb}). 
\\

{\bf Black Hole WIHs} $\Deltap$:\, Following the terminology used in the literature for black hole WIHs, we will now drop the bars and denote the space-time by $(M, g_{ab})$, the intrinsic metric on $\Deltap$ by $q_{ab}$, the intrinsic derivative operator by $D$. For the same reason, the equivalence class of preferred null normals will now be denoted by $[\ell^a]$ (in place of $[\bk^a]$) and the 1-form conjugate to any given $\ell^a \in [\ell^a]$ by $n_a$, so that $\ell^a n_a =-1$ (in place of  $\bk^a \bj_a = -1$). 

The universal structure and the symmetry group of $\Deltap$ is the same as that of a general WIH $\WIH$. In particular, the symmetry vector fields $\xi^a$ satisfy (\ref{symmetry}). Given a foliation $\vo ={\rm const}$ of $\Deltap$ where $\vo$ is an affine parameter of $\ello^a$, their explicit expressions are given by
\be \label{xi1} \xi^a \= \big((\mathring{\varpi} + \betao)\vo + \mathring{\mathfrak{s}})\big) \ello^a \,+\, \epsilono^{ab} \Do_b \chio\, + \, \qo^{ab} \Do_b \betao\, . \ee
Here $\mathring{\mathfrak{s}}(\theta,\phi)$ is a general function on the 2-sphere of null generators of $\Deltap$; each of $\chio(\theta,\varphi)$ and $\betao(\theta,\varphi)$ is a linear combination of first three spherical harmonics defined by $\qo_{ab}$; and  $\epsilono^{ab},\, \qo^{ab}$ are the inverses of the area 2-form and the metric on the $\vo = {\rm const}$ cross-sections, respectively. In this representation, $d^a := \mathring{\varpi}\vo \ello^a$ is a dilation,\,  $\mathring{\mathfrak{s}} \ello^a$ are supertranslations,\, $\epsilono^{ab} \Do_b \chio\,$ are rotations, and $\betao\,\vo \ell^o + \qo^{ab} \Do_b \betao$ are boosts. 

Next, let us use Einstein's equations to simplify the dynamical equation (\ref{Ddot1}). First, the Ricci tensor contribution to the right side vanishes. Second, only the Newman-Penrose component $\Psi_2$ of the Weyl tensor contributes and, on any WIH, it is completely determined by $q_{ab}$ and $D$ via $\Psi_2 = \f{1}{4}\, \mathcal{R} +  \f{i}{2} \epsilon^{ab} D_a \omega_b$, where $\mathcal{R}$ is the scalar curvature of $q_{ab}$. Therefore, (\ref{Ddot1}) becomes:
\be \label{Ddot2} \dot{D}_a n_b\, = \, D_{(a}\omega_{b)} + \omega_a \omega_b \,- \textstyle{\f{1}{4}}\, \mathcal{R}\, q_{ab}\, . 
\ee
Note that every field on the right side is Lie dragged by $\ell^a$; they are time-independent. Therefore the time dependence of $D$ is now tightly constrained: it is entirely determined by the triplet $(q_{ab},\omega_a, D_a n_b)$ on any 2-dimensional cross-section of $\Deltap$. In other words, there are no radiative degrees of freedom on $\Deltap$; it represents a black hole boundary in equilibrium.
\\

{\bf Asymptotic WIHs:} $\scrip$.\, Recall (e.g., from \cite{aa-yau}) that a physical space-time $(M,g_{ab})$ is \emph{asymptotically flat at null infinity} if there exists a Manifold $\hM$ with boundary $\scrip$ that is topologically $\mathbb{S}^2\times \mathbb{R}$, and equipped with a metric $\hg_{ab}$ such that $\hM = M\cup \scrip$ and $\h{g}_{ab} = \Omega^2 g_{ab}$ on $M$ where $\Omega$ is  a nowhere vanishing on $M$, and,\\
(i) $\Omega \= 0$, while \,$\h\nabla_a \Omega$\, is nowhere vanishing on $\scrip$; and,\\
(ii) The physical metric $g_{ab}$ satisfies vacuum Einstein's equations $R_{ab} = 0$ in a neighborhood of $\scrip$.

As is well known, the conformal Einstein equations satisfied by $\hg_{ab}$ imply that $\scrip$ is null and, without loss of generality, one can choose\, $\Omega$\, so that $\h\nabla^a \hn_a \=0$ where $\hn^a = \h\nabla^a \Omega$ is the null normal to $\scrip$. As is standard in the literature on null infinity, \emph{we will restrict ourselves to such divergence-free frames}.  Then conformal Einstein's equations imply $\h\nabla_a \hn^b \= 0$. Therefore integral curves of $\hn^a$ are affinely parametrized null geodesics with vanishing expansion and shear; $\scrip$ is a WIH. Furthermore, the 1-form $\omega_a$ vanishes identically, whence in particular its divergence vanishes. Thus $\hn^a\, \=\, \h\nabla^a \Omega$ serves as the \emph{preferred} normal to this WIH. Note that, while a generic WIH $\WIH$ is equipped with only an equivalence class $[\bk^a]$ of such null normals, the given conformal completion endows $\scrip$ with a specific preferred normal $\hn^a$; the freedom to rescale by a constant has disappeared.

As on $\Deltap$, let us adapt notation to the standard usage in the literature on $\scrip$. Thus, $(\bM, \bg_{ab})$ of general WIHS $\WIH$ will now replaced by a divergence-free completion $(\hM, \hg_{ab})$;\,  $(\bq_{ab}, \bk^a)$ by $(\h{q}_{ab}, \h{n}^a)$;\, and the 1-form $\bj_a$  will now be denoted by $\h\ell_a$ so that $\hn^a\,\h\ell_a = -1$. Of course, given a physical space-time, we can introduce another divergence-free conformal frame $\Omega^\prime = \mu  \Omega$ where $\mu$ is nowhere vanishing on $\hM$ and satisfies $\Lie_{\hn} \mu\=0$. In the WIH perspective, the resulting $(\hM, \hg_{ab})$ and  $(\hM, \hg^\prime_{ab})$ are to be regarded as distinct space-times, in both of which $\scrip$ is a WIH but with distinct WIH geometries $(\hq_{ab},\hn^a, \hD)$ and $(\hq_{ab}^\prime,\, \hn^{\prime\,a}, \hD^\prime)$. Put differently, while the fact that $\scrip$ is an extremal WIH is a property of the given physical space-time, the geometry of this WIH varies from one (divergence-free) conformal completion to another. The physics of this family of WIHs is the same; final results are all conformally invariant as usual. 

As in the case of $\Deltap$, let us begin with a discussion of the universal structure and symmetries of this WIH. They are essentially the same as on a general WIH \emph{but with a small variation}. As remarked above, while the given metric $\b{g}_{ab}$ endows geometrical WIHs $\WIH$ only with equivalence class $[\bk^a]$ of null normals, $\hg_{ab}$ endows $\scrip$ with a \emph{single} null normal $\hn^a$ through $\Omega$. Therefore, the universal structure that descends on $\scrip$ from the WIH perspective consists of pairs $(\qo_{ab}, \no^a)$, a unit round metric and a null normal, conformally related to $(\hq_{ab}, \hn^a)$ as on $\WIH$, so that any two pairs are related by $(\qo_{ab}^\prime, \no^{\prime\,a})\,=\,(\alphao^2\qo_{ab},\,\alphao^{-1}\no^a)$. In the more familiar terminology, these pairs correspond to Bondi conformal frames. Now, in coordinate-free discussions of $\scrip$, Bondi conformal frames play only a marginal role; one works directly with the pairs $(\hq_{ab}, \hn^a)$ (see, e.g., \cite{aa-yau}). In the WIH perspective, on the other hand, these pairs correspond to distinct WIH geometries; what is universally available on all of them are the pairs $(\qo_{ab}^\prime, \no^{\prime\,a})$ that correspond to Bondi conformal frames. Thus, when looked at through the WIH lens, symmetries are now those diffeomorphisms of $\scrip$ that map any one of these Bondi frames to another one. This subgroup is the BMS group $\B$ since $\B$ can also be characterized in this way.

For symmetry vector fields, the WIH perspective informs us that $\xi^a$ represents an infinitesimal symmetry if it satisfies Eq. (\ref{symmetry}), but now with $\mathring{\varpi} =0$ since there is no longer a freedom to rescale the normals $\no^a$ by constants. Thus in terms of (\ref{xi1}), we no longer have dilations that rescaled $\ko^a$ by constants, leaving the metric $\qo_{ab}$ unchanged. Given a pair $(\qo_{ab}, \no^a)$, then, the symmetry vector fields at $\scrip$ have the form:
\be \label{xi2}  \xi^a \= \big( \betao\,\uo + \mathring{\mathfrak{s}})\big) \no^a \,+\, \epsilono^{ab} \Do_b \chio\, + \, \qo^{ab} \Do_b \betao\, , \ee
where $\uo$ is now an affine parameter of $\no^a$. These are the standard expressions of BMS vector fields in a Bondi conformal frame $(\qo_{ab}, \no^a)$. Thus, the symmetry group $\G = \B \ltimes \mathfrak{D}$ of a general WIH $\WIH$ just `loses the dilation part' and reduces to the BMS group $\B$.

On the other hand, there is a strong contrast between the evolution equation of $\hD$ on $\scrip$ and that of $D$ on $\Deltap$: the Weyl tensor $\b{C}_{abc}{}^d$ as well as the 1-form $\b\omega_a$ on the right side of (\ref{Ddot1}) vanish on $\scrip$ and the evolution is governed by the Ricci part, which was the only one that vanished on $\Deltap$! Thus, in contrast to the Eq. (\ref{Ddot1}) on $\Deltap$  at $\scrip$ have:
\be \label{Ddot3}\dot{\hD}_a \h\ell_b\, = \textstyle{\f{1}{2}}\,\,\big(\h{S}_{\pb{ab}} + (\alpha\,- \textstyle{\f{1}{6}}\h{R})\, \hq_{ab}\big)\, . \ee
Now, the Bondi news is given by $\hN_{ab} = \h{S}_{ab} - \h\rho_{ab}$, where $\h\rho_{ab}$ is the (kinematical) Geroch tensor field \cite{gerochrev} that serves to remove the unphysical information contained in the conformal factor in $\hS_{\pb{ab}}$ (see, e.g., \cite{gerochrev,aa-radiativemodes,aa-yau}). Thus, the time-dependence of $\hD$ is dictated by the Bondi News that carries two degrees of freedom per point of 3-dimensional $\scrip$. To summarize, $\scrip$ is a WIH w.r.t. $\hg_{ab}$ that satisfies conformal Einstein's equations while $\Deltap$ is a WIH w.r.t. $g_{ab}$ that satisfies Einstein's equations. It is this difference that leads to diametrically opposite physics on the two WIHs, even though one starts from the same fundamental dynamical equation (\ref{Ddot1}). Thus, the unified WIH framework endows $\Deltap$ and $\scrip$ with essentially the same universal structures and symmetry groups and, at the same time, teases out diagonally opposite physics from them.\\

\textbf{Phase spaces of local DOF:} We will now summarize a new Hamiltonian framework, geared to local degrees of freedom, that serves as a common platform to discuss fluxes of physical observables associated with symmetries at both $\scrip$ as well as $\Delta^{\!{}^+}$. 

Recall that the standard covariant phase space $\ps$ consists of solutions $g_{ab}$ to field equations on the entire space-time $M$ satisfying the standard boundary conditions at $\scrip$ or/and $\Deltap$ (see, e.g., \cite{abr,wz,cfp,akkl2}). One can extract from it phase spaces $\psRh$ and $\psR$ that are adapted to open regions {\footnotesize{$\Rh$}} and {\footnotesize{$\R$}} of $\scrip$ and $\Deltap$, respectively, each bounded by arbitrary 2-sphere cross-sections \cite{aass1}. $\psRh$ consists of certain equivalence classes $\{\h{D}\}$ of connections on {\footnotesize{$\Rh$}} (where the equivalence relation serves just to get rid of a redundant conformal freedom). They encode precisely  the two radiative degrees of freedom of the gravitational field in full non-linear general relativity \emph{that reside in} {\footnotesize ${\h\R}$} \cite{aa-radiativemodes}. In particular, they do not capture any of the `Coulombic information' contained in $g_{ab}$. Similarly, $\psR$ consists of the (time independent) pair $(q_{ab},\,\omega_a)$, evaluated at a cross-section of $\Deltap$. These fields encode just the Coulombic information in $g_{ab}$; although there may be radiation  arbitrarily close to $\Deltap$, as in Robinson-Trautman space-times \cite{pc}, none of it registers in $\psR$! Thus, there is an interesting complementarity. 

In addition, $\psRh$ and $\psR$ have an extremely simple structure that greatly facilitates the subsequent analysis: while $\ps$ is a complicated non-linear space, $\psRh$ is an affine space, and $\psR$ is a convex set in a vector space. (This contrast can be traced back to the fact that elements of $\psRh$ and $\psR$ are freely specifiable; they are not subject to non-linear equations). Thanks to this `effective linearity', one can endow $\psRh$ and $\psR$ with a topology in which two elements are close to one another if they and their first derivatives on {\footnotesize{$\Rh$}} and {\footnotesize{$\R$}}  are close in the $L^2$-sense. This topology is motivated by detailed considerations involving test fields on asymptotically flat space-times \cite{aams,aass1}.

The symplectic current $\J_{abc}$ on $\ps$ is a 3-form whose integral over a Cauchy surface provides the symplectic structure $\bfomega$ of $\ps$. Integrals of $\J_{abc}$ on {\footnotesize{$\Rh$}} and {\footnotesize{$\R$}} provide symplectic structures $\bfomegaRh$ and $\bfomegaR$ on $\psRh$ and $\psR$ respectively. With this setup at hand, one can compute the Hamiltonians associated with symmetries. Let us begin with $\scrip$.  Then each BMS generator $\xi^a$ defines a Hamiltonian vector field $\delta_\xi$. However, as is common on infinite dimensional phase spaces (see, e.g., \cite{cm}), $\delta_\xi$ is only densely defined on $\psRh$. Therefore one first calculates the corresponding Hamiltonian $H_\xi (\{\hD\})$ on this subspace and asks if it can be continuously extended to the full phase space.%
\footnote{If one cannot, then one would have the `non-integrability' problem discussed in in other approaches, see e.g. \cite{wz,Barnich:2001jy,Barnich:2007bf,cfp,Freidel:2021cjp,Adami:2021nnf,Grant:2021sxk,Odak:2022ndm}.}
The answer is in the affirmative for all BMS vector fields $\xi^a$ and the result is: 
\be \label{H} H_\xi (\{\hD\}) = \f{1}{16\pi G}\,\,\int_{\scrip} \big[(\Lie_{\xi} \hD_a - \hD_a \Lie_{\xi})\h\ell_b + 2 \h\ell_{(a} \hD_{b)} \beta \big]\,\h{N}_{cd}\, \hq^{ac}\hq^{bd}\,\, \h\epsilon_{mnp}  \ee
where $\beta$ is given by $\Lie_{\xi} \hq_{ab} = 2\beta \hq_{ab}$ and $\h\epsilon_{mnp}$ is the volume 3-form on $\scrip$. This entire procedure can be applied to familiar, simple systems such as a massless scalar field in asymptotically flat space-times, and the Hamiltonians thus obtained agree with the standard expressions of BMS fluxes across 3-d regions {\footnotesize{$\Rh$}} of $\scrip$, obtained using the stress-energy tensor of the scalar field \cite{aams,aass1}. These applications serve to motivate the specific topology we use on $\psRh$ and lead one to interpret $H_\xi(\{\hD\})$ of (\ref{H}) as the flux $F_\xi [\Rh]$ of the BMS momentum across the region {\footnotesize{$\Rh$}} of $\scrip$. Indeed, the right side of (\ref{H}) is the standard expression of BMS fluxes, now obtained using the phase space $\psRh$ of degrees of freedom local to the region {\footnotesize{$\Rh$}} of $\scrip$. Note that  the procedure does not require additional structures such as preferred symplectic potentials or extensions of BMS vector fields away from $\scrip$.  Finally for open regions {\footnotesize{$\R$}} of $\Deltap$, one can apply exactly the same procedure but now the result is trivial: the pull-back of the symplectic current $\J_{abc}$ to $\Deltap$ vanishes identically \cite{cfp,akkl2}! (This is a reflection of the fact that there are no 3-d degrees of freedom in $\psR$.) Consequently, fluxes  $F_\xi$ {\footnotesize{$[\R]$}} also vanish identically for all generators $\xi^a$ of the $\Deltap$-symmetry group $\G$, just as one would expect physically. 

Thus, there is a single framework involving phase spaces of degrees of freedom that reside in open regions of WIHs. But it yields very different results depending on which of the two types of WIHs it is applied to. When applied to $\Deltap$, it predicts zero fluxes. When applied to $\scrip$, it yields the expressions of BMS fluxes with all the subtleties; e.g., including both the so-called `hard and soft terms' in the expression of the BMS supermomenta.

Finally, a natural extension of this framework leads to 2-surface charges $Q_\xi [S]$ associated with cross-section $S$ of $\scrip$ or $\Deltap$. The starting point is the observation that (conformal) Einstein's equations (and Bianchi identities) imply that the flux 3-forms --that constitute the integrand of the right side of (\ref{H})-- are exact, whence the integrals of their 2-forms potentials provide us with charges $Q_\xi [S]$.  However, this procedure requires us to go beyond the  phase spaces of local degrees of freedom  use additional inputs. In particular, one has to return to full $\ps$ since one needs to use full field equations, not just their pull-backs to the WIHs. In the case of $\scrip$, $\ps$ enables one to access the `Coulombic' degrees of freedom that are absent in $\psRh$ but are needed to obtain the explicit expressions of the charge 2-forms. These results, together with a comparison with other procedures \cite{wz,cfp} to define charges, appear in the companion paper \cite{aass1}.\\

\textbf{Discussion:} To summarize, $\Deltap$ and $\scrip$ share a number of properties, inherited directly from geometric WIHs $\WIH$. In particular, their intrinsic metrics are non-dynamical and their symmetry groups are almost the same. In both cases, the intrinsic connections $D$ and $\hD$ are dynamical, satisfying Eq.~(\ref{Ddot1}) that holds on any $\WIH$.  But complementary terms on the right side of this equation trivialize: While the Ricci terms vanish in $\Deltap$, the Weyl term vanishes on $\scrip$. Thanks to this subtle and surprising `complementarity', $\Deltap$ and $\scrip$ can  share  a large number of properties and, at the same time, display diametrically opposite physics. $\Deltap$ represents black hole horizons in equilibrium that lie in the strong field regime, and no radiation crosses them. $\scrip$ lies in the near-flat asymptotic regime and provides the arena for discussions of fluxes carried away by gravitational waves! This unified WIH framework can provide a foundation for `gravitational wave tomography' \cite{aa-banff}-- imaging horizon dynamics using black hole mergers using waveforms at $\scrip$-- and also have applications to the analysis of black hole evaporation in quantum gravity.

This interwoven structure is reminiscent of a musical fugue in four voices. The first voice is that of a geometric WIHs $\WIH$. It introduces the theme/subject. The second voice is that of black hole horizons $\Deltap$. It picks up on the theme and further develops it, providing us with one interpretation of the main subject that conveys the serenity of equilibrium. The third voice is that of null infinity $\scrip$. It continues the main theme but is contrapuntal because it exudes change and dynamism, with potential for bursts of energy and momentum. The fourth voice returns to the central idea enunciated by the first: a single theme --now, the phase space of local degrees of freedom-- can display the deep-down unity in spite of the diversity of manifestations.

\section*{Acknowledgments}

\noindent This work was supported in part by the Eberly and Atherton research funds of Penn State and the Distinguished Visiting Research Chair program of the Perimeter Institute. 

\providecommand{\href}[2]{#2}\begingroup\raggedright\endgroup

\end{document}